# Localized Oscillatory Dissipation in Magnetopause Reconnection

**J. L. Burch[1], R. E. Ergun[2], P. A. Cassak[3], J. M. Webster[4], R. B. Torbert[5,1], B. L. Giles[6], J. C. Dorelli[6], A. C. Rager[7,6], K.-J. Hwang[6], T. D. Phan[8], K. J. Genestreti[9], R. C. Allen[10], L.-J. Chen[11], S. Wang[11], D. Gershman[6], O. Le Contel[12], C. T. Russell[13], R. J. Strangeway[13], F. D. Wilder[2], D. B. Graham[14], M. Hesse[15], J. F. Drake[11], M. Swisdak[11], L. M. Price[11], M. A. Shay[16], P.-A. Lindqvist[17], C. J. Pollock[18], R. E. Denton[19], and D. L. Newman[2]**

[1]Southwest Research Institute, San Antonio, TX, USA.
[2]University of Colorado LASP, Boulder, CO, USA.
[3]West Virginia University, Morgantown, WV, USA.
[4]Department of Physics and Astronomy, Rice University, Houston, TX, USA.
[5]University of New Hampshire, Durham, NH, USA.
[6]NASA, Goddard Space Flight Center, Greenbelt, MD, USA.
[7]Catholic University of America, Washington, D. C., USA.
[8]University of California, Berkeley, CA, USA.
[9]Space Research Institute, Austrian Academy of Sciences, Graz, Austria.
[10]University of Texas, San Antonio, TX, USA.
[11]University of Maryland, College Park, MD, USA.
[12]Laboratoire de Physique des Plasmas, CNRS, Ecole Polytechnique, UPMC Univ Paris 06, Univ. Paris-Sud, Observatoire de Paris, Paris, France.
[13]University of California, Los Angeles, CA, USA.
[14]Swedish Institute of Space Physics, Uppsala, Sweden.
[15]Department of Physics and Technology, University of Bergen, Norway.
[16]University of Delaware, Newark, DE, USA.
[17]Royal Institute of Technology, Stockholm, Sweden.
[18]Denali Scientific, Healy, AK, USA.
[19]Dartmouth College, Hanover, NH.

Corresponding author: James Burch (jburch@swri.edu)

**Abstract**

Data from the NASA Magnetospheric Multiscale (MMS) mission are used to investigate asymmetric magnetic reconnection at the dayside boundary between the Earth's magnetosphere and the solar wind (the magnetopause). High-resolution measurements of plasmas, electric and magnetic fields, and waves are used to identify highly localized (~15 electron Debye lengths) standing wave structures with large electric-field amplitudes (up to 100 mV/m). These wave structures are associated with spatially oscillatory dissipation, which appears as alternatingly positive and negative values of $\mathbf{J} \cdot \mathbf{E}$ (dissipation). For small guide magnetic fields the wave structures occur in the electron stagnation region at the magnetosphere edge of the EDR. For larger guide fields the structures also occur near the reconnection x-line. This difference is explained in terms of channels for the out-of-plane current (agyrotropic electrons at the stagnation point and guide-field-aligned electrons at the x-line).

## 1 Introduction

Magnetic reconnection, a plasma process by which magnetic fields from different sources interconnect allowing plasma and momentum to be transferred across their boundaries, is important for plasmas in near and deep space and in the laboratory [*Burch and Drake*, 2009]. The process leading to reconnection involves explosive conversion of magnetic energy to heat and kinetic energy of charged particles. Here we show with data from the NASA Magnetospheric Multiscale (MMS) mission that the reconnection process at the dayside boundary of the Earth's magnetosphere is often driven by large electric-field components of highly confined wave-like structures with characteristics of oblique whistler waves. This result is very different from the lower-magnitude, widespread, quasi-static electric fields that are also found by observation and simulation to drive reconnection [*Shay et al.*, 2007; *Pritchett and Mozer*, 2009].

For asymmetric reconnection, with different plasma and magnetic pressures on either side of the boundary, the x-line (which separates magnetic field lines with different topologies) and the electron stagnation region (which marks the deepest penetration of magnetosheath electrons) separate along the Earth-Sun line [*Cassak and Shay* 2007]. For symmetric reconnection, as in the geomagnetic tail, these regions coincide. As shown in this letter, strong dissipation that is highly localized within the electron dissipation region (EDR) can occur at both the x-line and the electron stagnation region, and which region dominates depends on the out-of-plane magnetic-field component in the upstream region (the guide field). Particle-in-cell simulations of one of the same events by *Swisdak et al.* [2017] show the development of the standing oblique whistler waves and the strong dissipation they produce near the reconnection x-line and in the electron stagnation region. We note that localized strong dissipation has also been reported from simulations by *Zenitani et al.* [2011] and *Pritchett* [2013].

## 2 Observations

The NASA Magnetospheric Multiscale (MMS) mission extends the experimental investigation of magnetic reconnection in the boundary regions of the Earth's magnetosphere to the electron scale [*Burch et al.*, 2016a]. The present study investigates the structure of the electron dissipation region (EDR) for two reconnection events observed at the dayside magnetopause, one on October 16, 2015 (event 1) and the other on December 8, 2015 (event 2) [*Burch et al.*, 2016b; *Burch and Phan*, 2016]. Both of these previous studies investigated electron distribution functions, currents and electric fields in asymmetric reconnection, which is found at the dayside magnetopause. Asymmetry in magnetopause reconnection is caused by the high magnetic pressure and low plasma pressure on the magnetosphere side of the reconnection current layer and the lower magnetic pressure and higher plasma pressure on the magnetosheath side [*Cassak and Shay*, 2007]. Figure 1i shows a sketch of the typical magnetic reconnection geometry at the magnetopause with estimated trajectories of MMS for events 1 and 2.

Event 1 had a very small guide field (magnetic field component out of the page) of $B_M/B_L$ ~0.1 in boundary-normal coordinates [*Sonnerup and Cahill*, 1967; *Denton et al.*, 2016] while event 2 had a moderate guide field with $B_M/B_L$ ~ 1. A previous study [*Burch et al.*, 2016b] showed that in event 1 strong dissipation, as measured by $\mathbf{J} \cdot (\mathbf{E} + \mathbf{v}_e \times \mathbf{B}) > 0$, occurred in the electron stagnation region (S in Figure 1i) but very little dissipation occurred near the x-line (X in Figure 1i). It was shown further that the out-of-plane current associated with the dissipation ($J_M$) was carried by crescent-shaped electron distributions as had been predicted [*Hesse et al.*, 2014; *Chen et al.*, 2016].

In contrast, in event 2 it was found that significant out-of-plane currents occurred both near the x-line and in the electron stagnation region [*Burch and Phan*, 2016]. It was concluded that the guide field provided a channel for electron flow and the resulting out-of-plane current near the x-line that is not present for very small guide fields. Event 2 was further distinguished from event 1 in that trajectory of the spacecraft constellation was approximately normal to the magnetospheric boundary as shown in Figure 1i.

Here we examine with the highest resolution available the electric field, currents and electron distribution functions within the EDR for both events in order to determine the driving force for magnetic field dissipation and the resulting interconnection of magnetic fields and acceleration of particles. We find in both events that the dissipation is associated with highly confined intense standing wave structures that have certain characteristics that are consistent with oblique whistler waves.

**2.1 October 15, 2015 event.** Event 1 was the subject of the paper by *Burch et al.* [2016b], which identified the EDR in the electron stagnation region with electric field data averaged to the 30-ms time scale of the 3D electron distributions. Also identified by Burch et al. [2016b] were the crescent-shaped electron distribuion functions in the plane perpendicular to **B** and their evolution to parallel crescents during the transition from closed to open magnetic field lines. For the current study we analyzed the highest resolution DC eletric field data (8192/s) combined with 7.5-ms electron distribution functions described by *Rager et al.* [2017]. This higher-resolution analysis revealed the wave-like nature of the reconnection electric field and the out-of-plane current, which resulted in an oscillatory $\mathbf{J} \cdot \mathbf{E'}$ structure. We show in Figure 1a-f the magnetic field, electron velocity, electric field, dissipation, electric field power spectral density (PSD) and magnetic field PSD for a 450 ms time period beginning at 13:07:02.150 UT on October 16, 2015. Vector quantities in Figure 1 are plotted in boundary normal (LMN) coordinates with transformation matrix from GSE coordinates shown in the caption. Noted in panel a are the approximate locations of the electron stagnation region and the x line. Panel b shows the electron velocity moments and panel c the high-resolution electric field data. The wave-like nature of the electric-field, which was not shown by the 30-ms resolution data of *Burch et al.* [2016b], is clearly represented by the higher-resolution data. Panel d shows $\mathbf{J} \cdot \mathbf{E'}$ obtained by using 30-ms average **E** as in the Burch et al. paper, which also showed a large positive value and an adjacent but small negative value. Although $\mathbf{J} \cdot \mathbf{E'}$ is a scalar quantity, it is illustrative to show the separate contributions of the three components, which indicates that the dissipation was mainly associated with the out-of-plane current $J_M$ and the out-of-plane (or reconnection) electric field $E_M$. The wave spectrograms in panels e and f show the highest intensities near the dissipation peak with electromagnetic waves at low frequencies and broadband mostly electrostatic waves above 100 Hz with highest intensities below $F_{ce}$ (electron cyclotron frequency). Wave analysis shown later identifies these waves with the whistler mode.

Figure 1g and h show vector electric field measurements for MMS2 and MMS3 zoomed in to the 130 ms interval noted by the red bar between panels f and g. Based on the very similar electric field patterns in Figure 1g and h and the ~0.5 km separation in N between MMS2 and MMS3 shown in Figure 1j, we deduce that the current layer separating the magnetosphere from the magnetosheath moved Earthward at a velocity estimated to be ~30 km/s, causing the two spacecraft to traverse the electron stagnation region and the boundary between open and closed field lines in quick succession. This inward motion was accompanied by a faster southward motion causing the four MMS spacecraft (which were moving much slower at a few km/s) to follow the approximate trajectory shown in Figure 1i. This trajectory is consistent with that derived for this event by *Denton et al.* [2016] and *Hasegawa et al.* [2017].

Evident in the measurements in Figure 1g and h are mostly positive boundary-normal electric fields ($E_N$) and bipolar parallel ($E_L$) and out-of-plane ($E_M$) electric fields with nearly equal magnitudes as in an oblique electrostatic wave. This wave structure, which for MMS2 has an amplitude >80 mV/m, was accompanied by out-of-plane currents ($J_M$) of about $10^{-6}$ amp/m$^2$ carried by electrons with crescent-shaped distributions [*Hesse et al.*,

2014; *Burch et al.*, 2016b]. Thus the dissipation rate, $\mathbf{J} \cdot (\mathbf{E} + \mathbf{v_e} \times \mathbf{B})$, in the negative $E_M$ half of the wave structure was >80 nW/m$^3$ as is discussed further in connection with Figure 2. These E-fields are about two orders of magnitude greater than predicted reconnection electric fields [*Shay et al.*, 2016]. However, the facts that (1) they exist over a radial distance less than both the skin depth ($c/\omega_{pe}$) and the Larmor radius (both of which are a few km) and (2) they are bipolar in nature cause the average $E_M$ over these characteristic electron scales to match more closely the expectations. We also note that the $E_L$ and $E_M$ signals exhibit a bifurcation with temporal width approximately equal to the cyclotron period (4 ms in Figure 1h), which may imply that the waves are amplified by electrons that are trapped by the parallel electric field components [*Kellogg et al.*, 2010]. We suggest a similar amplification for the events observed in the October 16, 2015 MMS event. We note in Figure 1c that the wave structure appears to propagate into the open field line region toward the x-line.

Electron distribution functions shown in Figure 2 show that the magnetic field-line topology changed from closed to open over a 0.5-km structure containing the large oscillating electric fields. The polar plots of electron DFs in Figure 2a-c are accompanied by line plots for measurements within the black and red sectors noted in the polar plots. For all three times plotted the red plot in the top row, which is in the plane perpendicular to **B**, shows magnetosheath electrons mixed with magnetospheric electrons in crescent distributions. Rows b and c show magnetic field-aligned electrons. For time (1) fluxes along + and - B are nearly equal with a broad peak indicating more energetic magnetospheric electrons counterstreaming along the field. At time (3) the red and black curves are clearly different, showing a mixture of magnetosheath (red) and magnetospheric (black) electrons indicating open field lines.

Time (2) shows an intermediate case between (1) and (3). The polar DF plots at times (2) and (3) show clearly parallel crescent distributions as reported earlier by *Burch et al.* [2016b]. Figure 2d shows that the dissipation peaks at time (1) and that it results in the breaking and reconnection of magnetic field lines as the closed field lines convert to open field lines at times (2) and (3).

**2.2 December 8, 2015 event.** Figure 3 plots similar data from MMS2 for event 2. Figure 3a-f show that, in contrast to event 1, for this moderate guide-field case there were large electron velocities in the M direction both near X and S, as noted before [*Burch and Phan*, 2016; *Genestreti et al.*, 2017]. Figure 3g-j shows 40-ms zoom data for the time denoted by the red bar below panel f. In this case $v_e$ is plotted at 7.5 ms time resolution [*Rager et al.*, 2017], showing a peak coincident with the strong electric-field signal in panel h, which has similar characteristics to that shown in Figure 1g-h with a positive $B_N$ signal accompanied by bipolar L and M components. Panel i shows the reversal of $B_L$ near a minimum of $B_N$, which is indicative of an in-plane null and x-line. The green curve in panel i shows the guide field. Panel j shows the dissipation while panels k and l show the effects on the electron DFs. Both before and after the oscillating electric field structure or standing wave (at 0.437 - 0.440 s) the magnetic field lines are open, as indicated by the off-scale (yellow) extension in the $-v_\parallel$ direction. This extension is from magnetospheric electrons resulting from the connection of the field line to the northern hemisphere. The lower-energy red regions in the polar DF plots indicate magnetosheath electrons moving along $+v_\parallel$. Near 0.440 s, centered on the whistler wave structure, there is a red area along $-v_\parallel$ in panel l, indicating an accelerated electron beam along the magnetic field, which is along the guide field. At this point, the reconnection electric field is $E_{-M}$ and the out-of-plane current is $J_{-M}$. As denoted in the green trace of panel j, there is dissipation leading to x-line reconnection at this point. The fact that there are open field lines on either side of the x-line is consistent with the diagram in Figure 1 with the spacecraft passing just north of the x-line. Referring to the middle plot in panel l, the strong electron beam along $-v_\parallel$, being more energetic than the opposite beam along $+v_\parallel$, is responsible for carrying the out-of-plane current and results from the reconnection electric field, which is mostly $E_{par}$ in this region. These conditions are consistent with current ideas about reconnection except for the important difference that they occur in a very restricted region of the EDR (smaller than the skin depth and the electron gyroradius) and involve electric fields two orders of magnitude larger than predicted.

**3 Wave Analysis**

An important issue to address is the nature and origin of the localized large oscillatory electric fields in the electron diffusion region. The observation of similar wave forms with decreasing amplitudes downstream of the intense events observed by both MMS2 and MMS3 on October 16, 2015 (Figure 1g,h) and by MMS2 on December 8, 2015 (Figure 3h) could be interpreted as the propagation of waves from the electron stagnation region into the exhaust region. This observation could also be interpreted as the encounter of residual spatial structures left behind by the Earthward motion of the magnetopause and reconnection current layer on October 16, 2015 and their Sunward motion on December 8, 2015.

In order to shed more light on the nature of the oscillatory electric fields and the associated magnetic field fluctuations, wave analysis has been performed for event 1 (Figure 4) and event 2 (Figure 5). In both cases within the frequency range containing the major wave structure obliquely propagating waves are determined by Poynting flux analysis, and hodograms show elliptical waves with right-hand polarization. In addition to the major structures, where for event 1 the duration is about 20 ms and for event 2 the duration is about 3 ms, there is superimposed higher frequency oscillations with much smaller amplitudes. In order to eliminate these higher frequency signals from the analysis we limited the analysis to two lower frequency ranges for each event. For event 1 the more relevant frequency range shown in Figure 4 is 10 - 50 Hz, and in this range there is clear right-hand (RH) polarization and a wave angle of about 30° based on the Poynting vector. In the higher-frequency range (100 - 300 Hz) the polarization trends toward linear. The Poynting-vector plot for frequencies between 10 and 50 Hz in Figure 4 shows a reversal of direction for all three components. Such a reversal indicates either a wave reflection or a source region.

For event 2 (Figure 5) a higher frequency range (200 - 500 Hz) is more relevant since the event lasted only ~3 ms. Again right-hand elliptical polarization is evident while the polarization becomes left-handed at the lower frequency range (10 - 100 Hz). As in event 1 there is an indication of a reversal of the Poynting vector, particularly for the field-aligned component in the 200 - 500 Hz frequency range. We conclude for both events that the wave analysis is consistent with oblique whistler mode oscillations although not conclusively so because of the short duration of the major oscillations.

We investigated the observed electron distribution functions as possible sources of the large oblique whistler-like oscillations with inconclusive results. We did find that the perpendicular crescent distributions as shown in Figure 2a are responsible for the mostly electrostatic waves observed between $F_{ce}$ and $F_{uh}$ (upper hybrid frequency) in Figure 1f, as has been shown previously by *Graham et al.* [2017] to result from a beam-mode interaction for an event on December 14, 2015. However, the parallel beams shown in Figures 2c and 3l were not found to generate oblique whistler-mode waves as might be expected from the relatively high beta (>5) in the two events [e. g., *Sauer and Sydora*, 2010]. Another consideration is the small width of the events, which are only 15 - 20 Debye lengths and only marginally larger than the theoretical limit of $2\pi\lambda_D$ for the shortest wavelength that can occur in a plasma. Thus it is possible that these are Debye-scale solitary structures rather than standing oblique whistler waves [e. g., *Ergun et al.*, 1998], and more analysis is clearly needed to determine definitively the nature and cause of these large oscillatory structures, which are associated with conversion of electromagnetic energy to particle energy in the electron diffusion region.

## 5 Conclusions

We have shown with MMS data that reconnection dissipation at the Earth's dayside magnetopause occurs in highly localized regions (~0.25 electron inertial lengths or ~15 electron Debye lengths) within the electron diffusion region. For an event with a very small guide magnetic field the only significant dissipation was located near the electron stagnation region, while for an event with guide field ~1 significant dissipation occurred both near the X-line and near the stagnation region. The general result that dissipation near the X-line depends on significant guide field has been described by *Genestreti et al.* [2017]. We have shown further that the dissipation is associated with an oscillatory electric-field pattern that shows characteristics of both a spatial structure and a propagating wave. We showed by wave analysis that the events are most likely standing oblique whistler waves. The oscillating electric fields, combined with fairly uniform out-of-plane currents, lead to alternating positive and negative J • E values with the positive values associated with the conversion of electromagnetic energy to particle energy and the negative values indicating conversion of particle energy to electromagnetic energy. When the electric field is averaged over the 30 ms electron distribution function sampling rate of MMS, as was done by *Burch et al.* [2016b], the J • E values are mostly positive. Further, by using the 7.5 ms electron distributions derived by *Rager et al.* [2017] we were able to show for the electron stagnation region of the October 16, 2015 event that the positive J • E peak corresponded exactly with the conversion of closed to open magnetic field lines.

*Swisdak et al.* [2017] show the development of structures they identify as standing oblique whistler waves at the electron stagnation region for parameters derived from the October 16, 2015 event we have described.

In their simulation the standing oscillatory structure is generated by electrons streaming along the outer reconnection separatrix that are accelerated through the magnetic null region by a strong $E_N$ component toward the electron stagnation regions. They find both positive and negative values of J • E associated with electric-field amplitudes up to 25 mV. Within the limitations of the PIC simulation, these results are consistent with the MMS observations.

Acknowledgments, Samples, and Data

This work was supported by NASA Contract No. NNG04EB99C at SwRI.

The entire MMS data set is available on-line at https://lasp.colorado.edu/mms/sdc/public/links/. Fully calibrated data are placed on-line at this site within 30 days of their transmission to the MMS Science Operations Center. The data are archived in the NASA Common Data Format (CDF) and so can be plotted using a number of different data display software packages that can use CDF files. A very comprehensive system called the Space Physics Environment Data Analysis System (SPEDAS) is available by downloading http://themis.ssl.berkeley.edu/socware/bleeding_edge/ and selecting spdsw_latest.zip. Training sessions on the use of SPEDAS are held on a regular basis at space physics related scientific meetings. All of the data plots in this paper were generated with SPEDAS software applied to the publicly available MMS data base, so they can readily be duplicated.

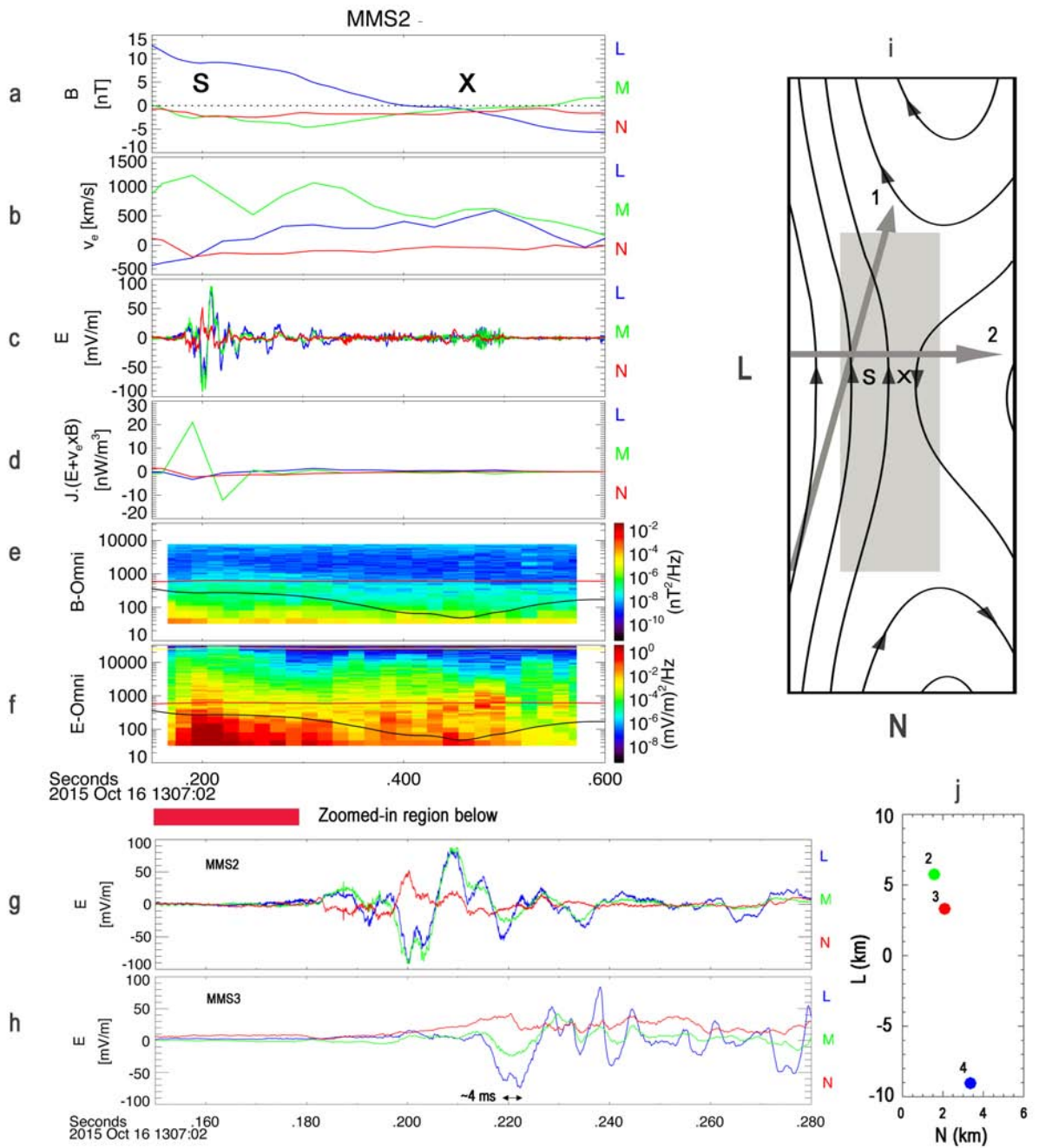

**Figure 1.** Electric and magnetic field and electron data for a reconnection event at the Earth's magnetopause on October 16, 2015. **a**, Vector magnetic field with approximate S (stagnation region) and X (x-line) noted. **b**, Electron velocity. **c**, Vector electric field at 8192/s. **d**, dissipation in plasma rest frame {**J** • (**E** + **v$_e$** x **B**)} averaged to the 30-ms electron measurement cadence. **e**, Electric power spectral density (PSD) with F$_{ce}$ (black curve) and F$_{pi}$ (ion plasma frequency, red curve). **f**, Magnetic PSD. **g**, Zoomed-in vector electric field from MMS2 at the 8192/s E-field measurement cadence. **h**, Same for MMS3. **i**, Sketch of magnetic field lines for asymmetric reconnection with shaded region for EDR, grey arrow for spacecraft trajectories for events 1 and 2 with S and X denoting electron stagnation region and reconnection x-line, respectively. **j**, Positions of MMS2, 3 and 4 for zoomed-in data period. All plots are in boundary-normal coordinates with transformation matrices from GSE to LMN coordnates: L = [ 0.31147, 0.02399, 0.94998]GSE, M= [ 0.48027, -0.8652, -0.13562]GSE, N= [ 0.81863, 0.49849, -0.28099]GSE [*Denton et al*., 2016].

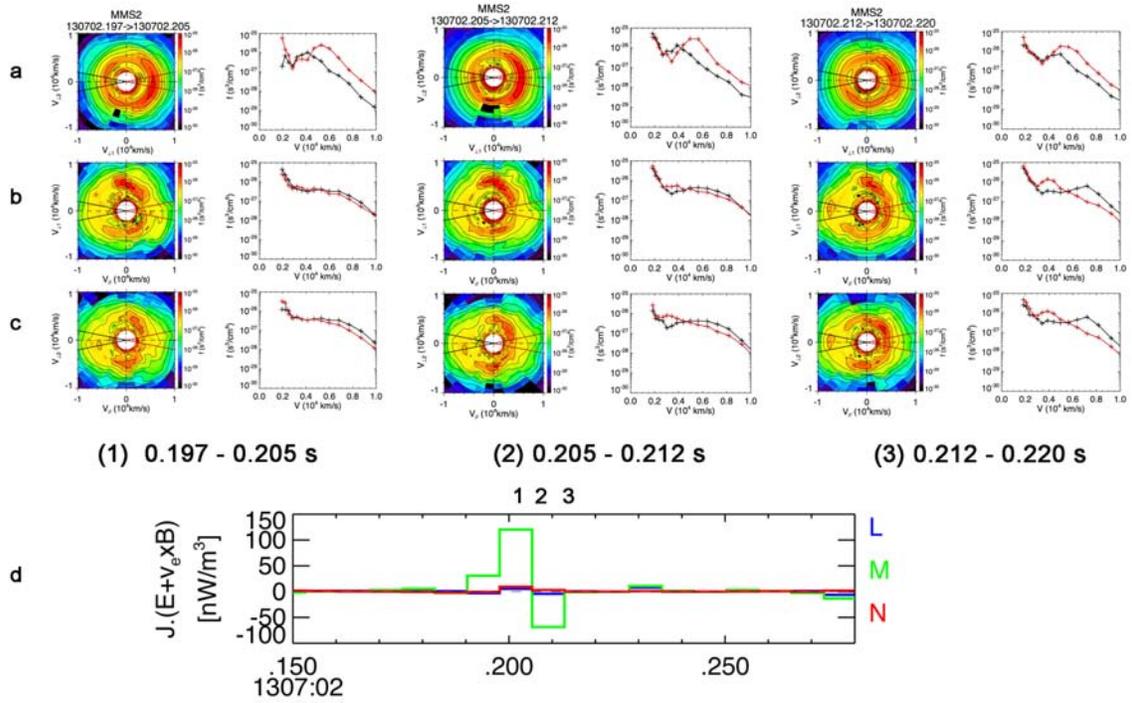

(1) 0.197 - 0.205 s        (2) 0.205 - 0.212 s        (3) 0.212 - 0.220 s

**Figure 2.** Electron distribution functions and dissipation every 7.5 ms for a 130 ms period on October 16, 2015. a, Electron distribution functions every 7.5 ms in the plane perpendicular to **B**. $V_{\perp 1}$ is in the $(\mathbf{b} \times \mathbf{v}) \times \mathbf{b}$ direction, which is a proxy for $\mathbf{E} \times \mathbf{B}$. $V_{\perp 2}$ is in the **E** direction. b, Electron distribution functions in the plane containing **B** and $V_{\perp 1}$. c, Electron distribution functions in the plane containing **B** and $V_{\perp 2}$. Line plots in b and c show average distribution function within the red and black sectors in each polar plot. d, dissipation in plasma rest frame $\{J \bullet (E + v_e \times B)\}$ every 7.5 ms.

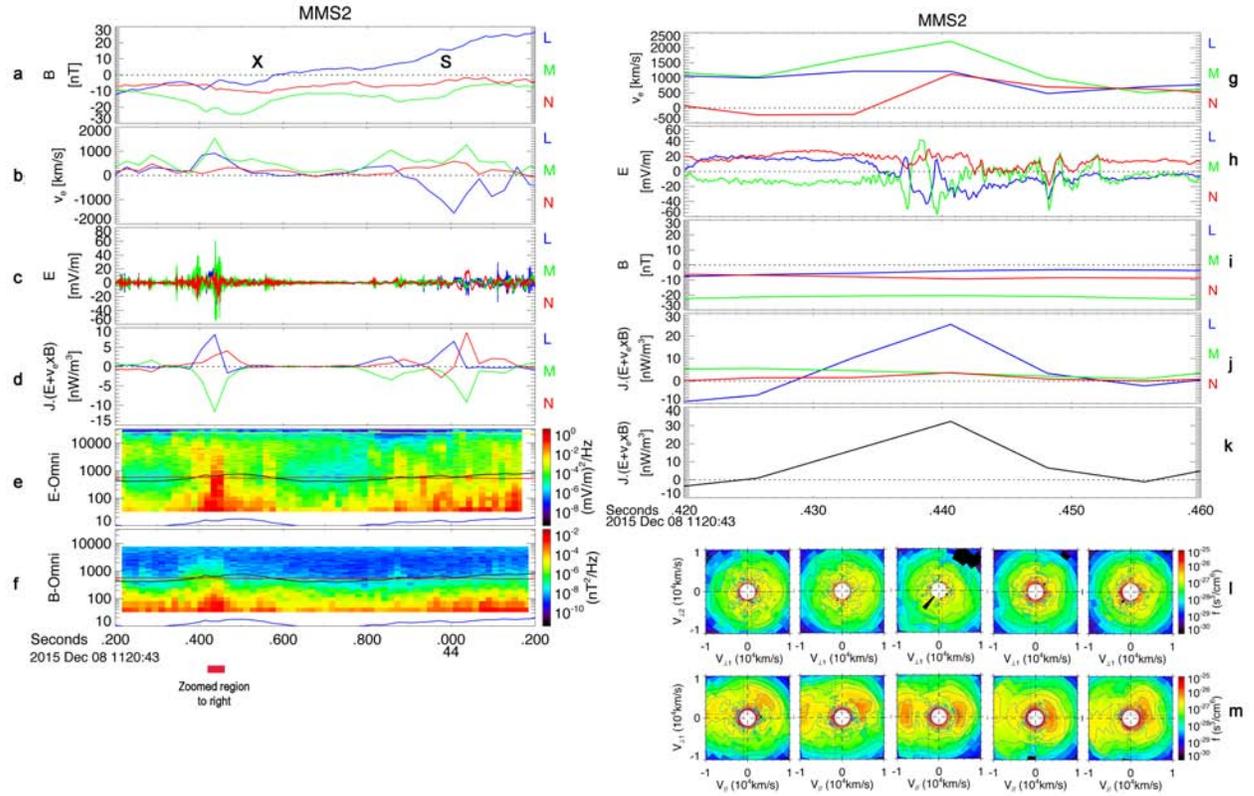

**Figure 3.** (left) Electric and magnetic field and electron data in a one second period on December 8, 2015. Format is the same as in Figure 1a-f. a, Vector magnetic field. b, Electron velocity every 30 ms. c, Vector electric field. d, dissipation in plasma rest frame {J • (E + v$_e$ x B)}. e, Electric power spectral density (PSD) with F$_{lh}$ (blue curve). f, Magnetic PSD. (right) data zoomed in to the 40 ms period shown by the red bar at bottom of left panel. g, Electron velocity every 7.5 ms. h, Vector electric field. i, Vector magnetic field. j, J • (E + v$_e$ x B) at 7.5-ms resolution showing L, M. and N separately. k, total J • (E + v$_e$ x B) at 7.5-ms resolution. l, Electron distribution functions every 7.5 ms in the plane perpendicular to B at times corresponding to the center of each plot. V$_{\perp1}$ is in the (**b** × **v**) × **b** direction, which is a proxy for **E** × **B**. V$_{\perp2}$ is in the **E** direction. m, Electron distribution functions every 7.5 ms in the plane containing **B** and V$_{\perp1}$. All line plots are in boundary-normal coordinates with transformation matrices from GSE to LMN coordnates: L = [0.30793923, -0.20121783, 0.92988430]GSE, M= [-0.010207845, -0.97802154, -0.20825384]GSE, N= [0.95135126, 0.054637414, -0.30322520]GSE.

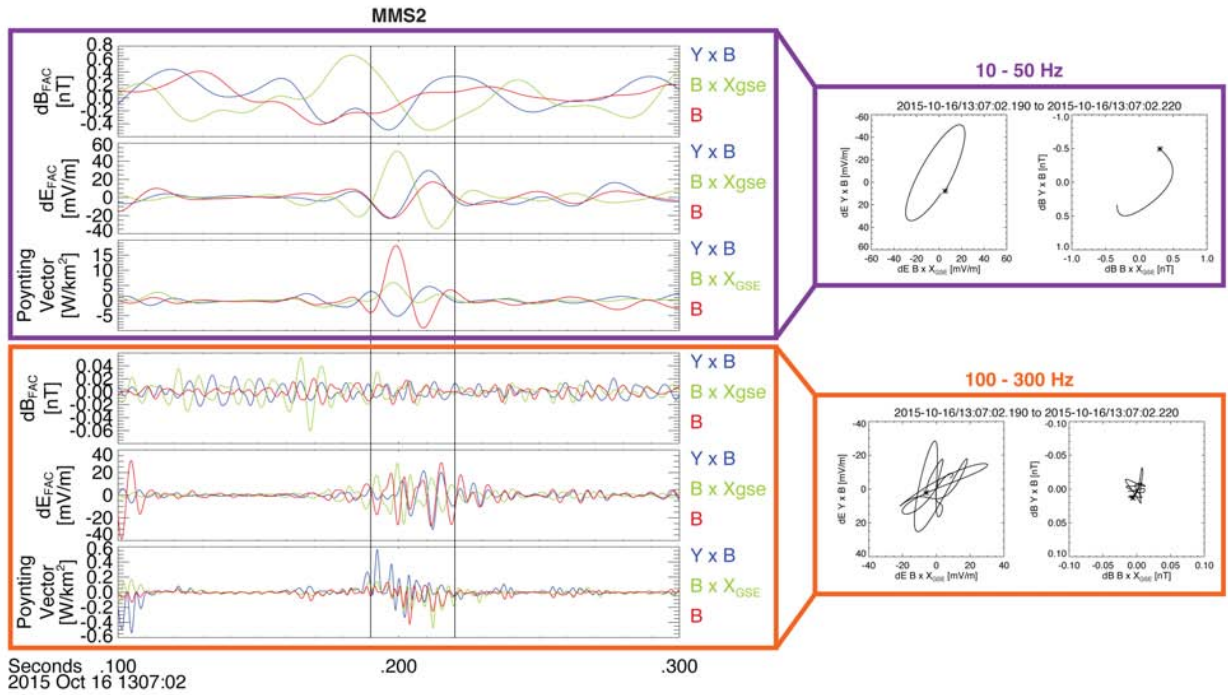

**Figure 4.** Wave analysis for event 1 on October 16, 2015. Data are in field-aligned coordinates. Top plot is for frequencies between 10 and 50 Hz. Bottom plot is for frequencies between 100 and 300 Hz. The magnetic field vector points out of the hodogram figures on the right.

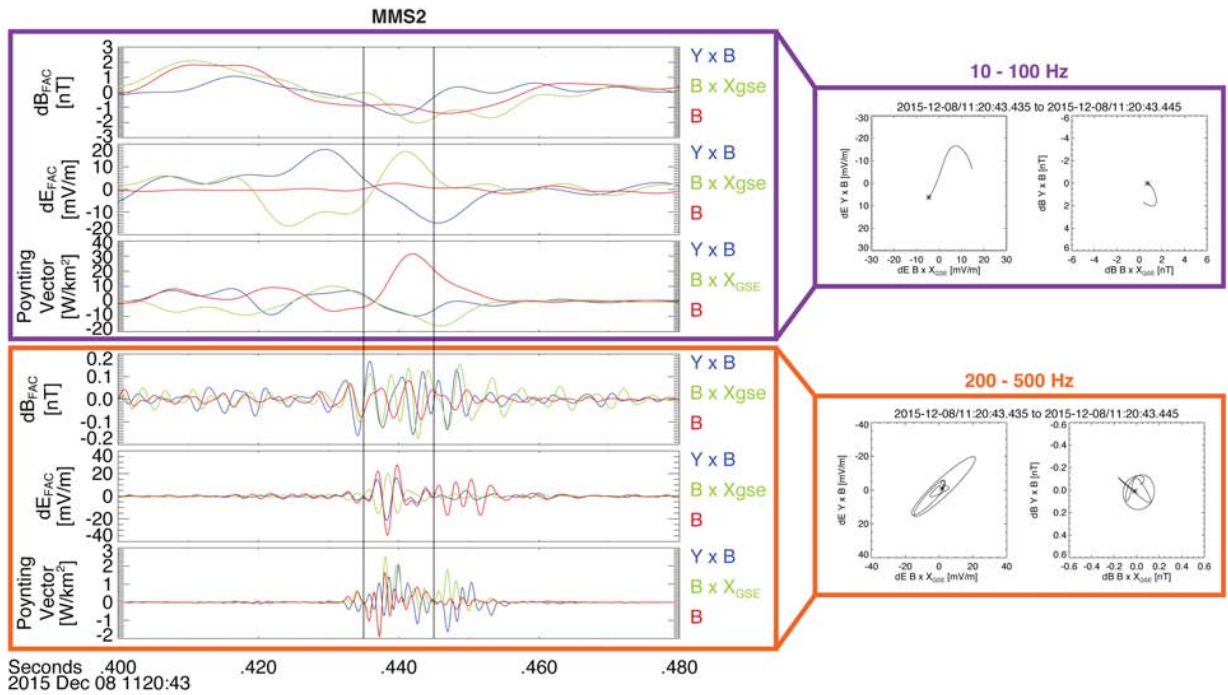

Figure 5. Wave analysis for event 2 on December 8, 2015. Data are in field-aligned coordinates. Top plot is for frequencies between 10 and 100 Hz. Bottom plot is for frequencies between 200 and 500 Hz. The magnetic field vector points out of the hodogram figures on the right.